\documentclass{jfm}%[referee]

\usepackage[T1]{fontenc}

\usepackage{amssymb}
\usepackage{amsmath}
\usepackage{graphicx}
\usepackage{tabularx}
\usepackage{cite}
\usepackage{color}
\usepackage{subfigure}
\usepackage{subfigmat}
\usepackage{etoolbox}
\usepackage{url}
\usepackage{bm}
\usepackage{siunitx}

\newcommand{\ve}[1]{\mathbf{#1}}
\newcommand{\bu}{\ve{u}}

\newcommand{\bI}{\ve{I}}

\title{Near-surface dynamics of a gas bubble collapsing above a crevice}

\author{
    Theresa Trummler\aff{1}\corresp{\email{theresa.trummler@tum.de}},
	Spencer H. Bryngelson\aff{2},
	Kevin Schmidmayer\aff{2},
    Steffen J. Schmidt\aff{1},
    Tim Colonius\aff{2} \and
    Nikolaus A. Adams\aff{1}
	}

\affiliation{
    \aff{1} Chair of Aerodynamics and Fluid Mechanics, Technical University of Munich \\ 
    Boltzmannstr.\ 15 85748 Garching bei M\"unchen, Germany
    \aff{2} Division of Engineering and Applied Science, California Institute of Technology \\  
    1200 E.\ California Blvd., Pasadena, CA 91125, USA
	}

\begin{document}

\maketitle

\begin{abstract}
    The impact of a collapsing gas bubble above rigid, notched walls is considered. Such surface crevices and imperfections often function as bubble nucleation sites, and thus have a direct relation to cavitation-induced erosion and damage structures. A generic configuration is investigated numerically using a second-order-accurate compressible multi-component flow solver in a two-dimensional axisymmetric coordinate system. Results show that the crevice geometry has a significant effect on the collapse dynamics, jet formation, subsequent wave dynamics, and interactions. The wall-pressure distribution associated with erosion potential is a direct consequence of development and intensity of these flow phenomena.
\end{abstract}

\maketitle

\section{Introduction}
\label{sec:Intro}

Cavitation damage can result from the collapse of vapour bubbles formed in low pressure regions of a flow, typically at gas nuclei that exist in the free-stream or in crevices on surfaces. When collapse occurs near a surface, the emitted shock waves ~\citep{rayleigh1917,hickling64} impinge on nearby surfaces~\citep{benjamin66,Plesset:1971hu} where, depending on the surface geometry and material properties, they can cause erosion or ablation.
The importance of surface geometry, in conjunction with the fact that these low-pressure regions occur more frequently at rough surfaces, motivates the investigation of bubble collapse behaviour near solid walls. 

\begin{figure}
    \centering
    \includegraphics[scale=1]{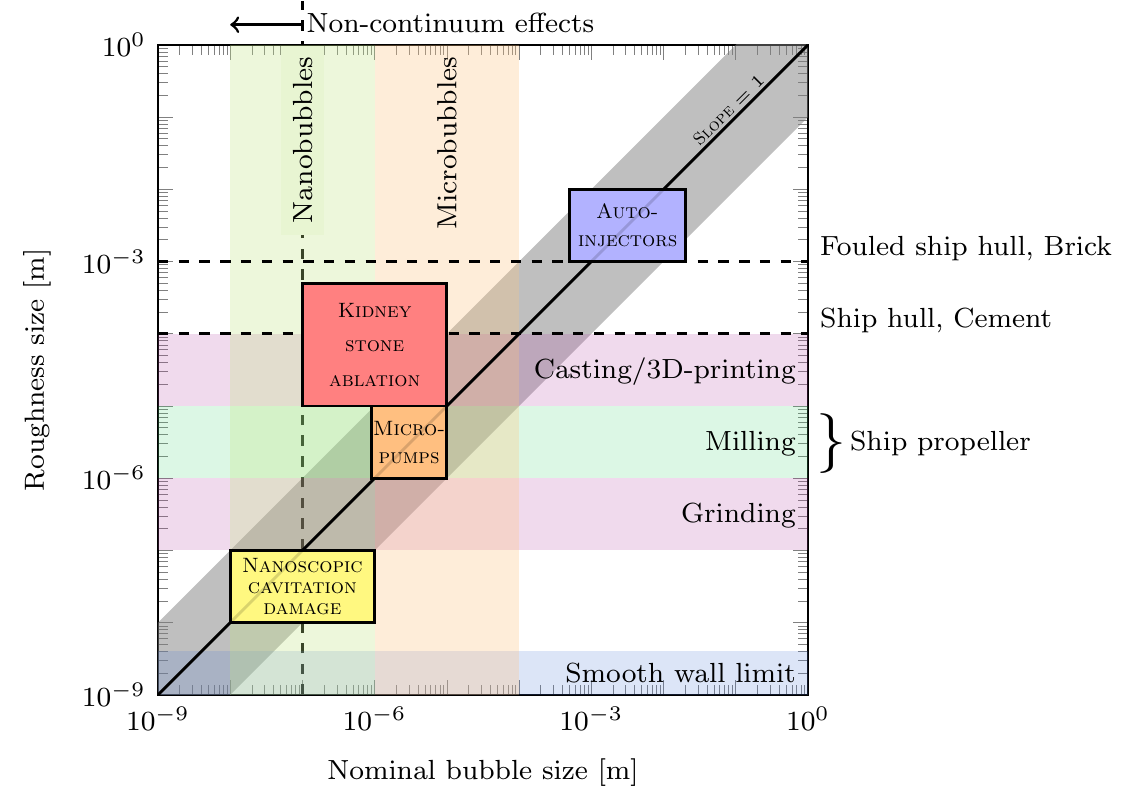}
    \caption{
        Nominal bubble size and surface roughness of common engineering materials, finishing processes, and applications.
        Roughness sizes are RMS values. 
        }
    \label{f:scales}
\end{figure}

Several studies have analysed bubbles collapsing near smooth walls.
Early studies identified an asymmetric behaviour associated with bubble-wall interaction~\citep{benjamin66,Plesset:1971hu} that leads to an impinging jet. Later, experimental studies have analysed the collapse behaviour~\citep{Lindau:2003oct}, jet formation and velocities~\citep{Tomita:1986gy}, and wall erosion potential~\citep{Philipp:1998eg} in greater detail. Numerical-simulation-based studies have investigated the collapse dynamics of bubbles attached to~\citep{Lauer:2012jh} and near~\citep{Johnsen:2009cua} smooth walls. However, such configurations represent the wall pressure and collapse dynamics only if the length scale of the wall roughness is much smaller than the nominal bubble size. When this condition is not satisfied, the bubble collapse, and thus its effect on near-wall erosion, can change qualitatively~\citep{Tomita:2002dn,Li:2018cx,Zhang:2018cav}. In figure~\ref{f:scales} physical limits and regions of relevant manufacturing processes and engineering applications are shown for a range of roughness sizes and bubble length scales. The associated broad list of applications, including urinary stones ablation~\citep{Pishchalnikov:2003xv}, surface cleaning~\citep{Ohl:2006fs,Reuter:2017bu}, cavitation in micro-pumps~\citep{dijkink2008laser} and pressurized auto-injectors~\citep{Veilleux:2018gy} and due to nano-bubbles~\citep{borkent2009nucleation} motivates the study of bubble collapse dynamics in this regime.

Our goal is to determine how a surface crevice modifies the collapse of a near-wall bubble, and to assess thus the associated modification of wall pressure, jet and shock formation, and wave interactions, which are of principal importance when considering erosion and damage potential~\citep{brennen95,Pohl:2015keb}. For this purpose, the collapse of a spherical gas bubble near or attached to a wall with a cylindrical notch is analysed. Experimental techniques preclude detailed visualization of such small space- and time-scale dynamics, particularly with respect to the rapid liquid jet formation and the high pressures waves emitted after collapse. Therefore, we use numerical simulations to characterize qualitative and quantitative differences of collapse behaviour associated with the surface geometry. 

In section~\ref{sec:Methods}, we describe the physical model and numerical method. The specific configurations considered are presented in section~\ref{sec:NumSetup}, and include variations in notch size and bubble-wall stand-off distance. The variation in notch size serves as a representation of the varying degrees of surface roughness present in engineering applications (see figure~\ref{f:scales}), whereas the stand-off distance has a significant impact on the collapse dynamics and wall-pressure for smooth-wall cases~\citep{Tomita:1986gy,Philipp:1998eg,Lauer:2012jh}. The collapse behaviour of the bubble is analysed for such configurations in section~\ref{sec:Results}, followed by a consideration of the collapse and jet-impact times, velocities, and wall-pressures. Section~\ref{sec:Conclusion} concludes the paper. 

\section{Physical model and numerical methods}
\label{sec:MathPhysModel}

\subsection{Governing equations}

The collapse of a gas bubble in liquid is modelled using a 6-equation multi-component flow model~\citep{Saurel:2009si} that conserves mass, momentum, and total energy. For the driving pressures and gas bubbles we assume that the effects of viscosity and surface tension are insignificant with respect to inertial effects, and so they are not included in our model. The governing equations are
\begin{align} 
\begin{split}
    \frac{\partial \alpha_l}{\partial t} + \bu \cdot \nabla \alpha_l &= \mu (p_l - p_g) , \\
    \frac{\partial \alpha_l \rho_l}{\partial t} + \nabla \cdot (\alpha_l \rho_l \bu) &= 0 , \\
    \frac{\partial \alpha_g \rho_g}{\partial t} + \nabla \cdot (\alpha_g \rho_g \bu) &= 0 , \\
    \frac{\partial \rho \bu }{\partial t} + \nabla \cdot (\rho \bu \bu + p \bI) &= \mathbf{0} , \\
    \frac{\partial \alpha_l \rho_l e_l}{\partial t} + \nabla \cdot (\alpha_l \rho_l e_l \bu) + \alpha_l p_l \nabla \cdot \bu &= - \mu p_I (p_l - p_g) , \\
    \frac{\partial \alpha_g \rho_l e_g}{\partial t} + \nabla \cdot (\alpha_g \rho_g e_g \bu) + \alpha_g p_g \nabla \cdot \bu &= \phantom{-} \mu p_I (p_l - p_g) ,
    \label{eq_general}
\end{split}
\end{align}
where $\alpha_k$, $\rho_k$, $p_k$ and $e_k$ are the volume fraction, density, pressure and internal energy of phase $k$, respectively, $\rho$, $p$, and $\bu$, are the mixture properties, $\mu$ is the relaxation coefficient, and $p_I$ is the interfacial pressure~\citep{Saurel:2009si}. The mixture total energy is
\begin{gather}
    E = e + \frac{1}{2} \| \bu \|^2,
\end{gather} 
where $e$ is the mixture specific internal energy
\begin{gather}
    e = \sum_{k=1}^2 Y_k e_k \left( \rho_k , p_k \right).
    \label{eq_internalEnergy}
\end{gather}
In~\eqref{eq_internalEnergy}, $e_k$ is defined via an equation of state and $Y_k$ are the mass fractions
\begin{gather}
    Y_k = \frac{\alpha_k \rho_k}{\rho} .
\end{gather}
The gas $g$ is modelled by the ideal-gas equation of state
\begin{gather}
    p_g = ( \gamma_g - 1) \rho_g e_g ,
\end{gather}
and the liquid $l$ is modelled by the stiffened-gas equation of state
\begin{gather}
    p_l = ( \gamma_l - 1) \rho_l e_l - 
    \gamma_l \pi_\infty,
\end{gather}
where $\gamma_g = 1.4$, $\gamma_l = 2.35$, and $\pi_\infty = \SI{e9}{\pascal}$ are model parameters~\citep{LeMetayer:2005gu}.

\subsection{Numerical method}
\label{sec:Methods}

A second-order-accurate MUSCL scheme is used to solve~\eqref{eq_general}. It is implemented in ECOGEN~\citep{schmidmayer2019ecogen,schmidmayer2019AMR}, which has been verified for several gas bubble dynamics problems, including free-space~\citep{schmidmayer19} and wall-attached~\citep{Pishchalnikov:2018pp} bubble collapses. The approach uses piece-wise linear reconstruction~\citep{toro97} of the primitive variables to suppress spurious oscillations at material interfaces~\citep{coralic2014WENO5}. The monotonized central (MC)~\citep{van1977MC} slope limiter and THINC interface-sharpening technique~\citep{shyue2014thinc} are used to minimize interface diffusion. The associated Riemann problem is computed using the HLLC approximate solver~\citep{Saurel:2009si, toro97}. An explicit two-step time integrator is used~\citep{schmidmayer2019ecogen}.

The pressure-non-equilibrium model~\eqref{eq_general} also requires pressure-relaxation to recover a unique equilibrium pressure. This is achieved by an infinite-relaxation procedure~\citep{Saurel:2009si}. At each time step it solves the non-relaxed, hyperbolic equations ($\mu \to 0$), then relaxes the non-equilibrium pressures for $\mu \to +\infty$. The relaxation procedure is combined with a re-initialization procedure at each time-step stage, which ensures a unique pressure and the conservation of total energy, and thus convergence to the 5-equation mechanical-equilibrium model~\citep{kapila2001}.

\section{Problem set-up}
\label{sec:NumSetup}

\begin{figure}
    \centering
    \includegraphics[scale=1]{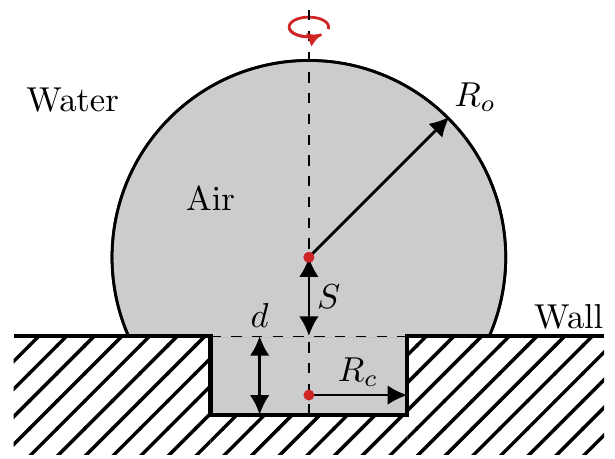} 
    \caption{Schematic of the problem set-up.} 
    \label{fig:sketch}
\end{figure}

Figure~\ref{fig:sketch} shows the flow configuration considered. The initial bubble is spherical with radius $R_0$ and stand-off distance $S$ above a cylindrical crevice of radius $R_C$ and depth $d = 0.25 R_0$. We define the stand-off distance $S$ as the distance from the wall to bubble-centre for $R_C/R_0 \leq 0.5$ and as the distance from the crevice-bottom to bubble-centre for $R_C/R_0 > 0.5$. This definition ensures consistency for both limiting cases $R_C/R_0 \to 0$ and $R_C/R_0 \to \infty$. 

We consider a $R_0 = \SI{400}{\micro\meter}$ bubble filled with non-condensable gas of initial pressure $p_B = \SI{3000}{\pascal}$ and density $\rho_g = \SI{0.03565}{\kilo\gram\per\meter\cubed}$. Bubbles commonly used in relevant applications predominately consist of non-condensable gas. Furthermore, the collapse dynamics are also only weakly sensitive to the internal bubble pressure when the driving pressure differences are large~\citep{Pishchalnikov:2018pp}.

The bubble is surrounded by water with a density of $\rho_l = \SI{1002.7}{\kilo\gram\per\meter\cubed}$ and varying pressure
\begin{gather}
    p(\hat{r},t=0) = p_{\infty} + \frac{R_0}{\hat{r}} \left( p_B - p_\infty \right)
    \;\; \text{for} \;\; \hat{r} > R_{0},
    \label{eqn:p_liq}
\end{gather}
where $\hat{r}$ is the radial coordinate with origin at the bubble centre. This initialization matches the pressure distribution predicted by the Rayleigh equation for the Besant problem~\citep{brennen95,besant59}. For the configurations considered, it provides a suitable approximation of the realistically evolving pressure field and suppresses the formation of spurious pressure waves due to pressure jumps. We use $p_\infty = \SI{e7}{\pascal}$, which matches that of previous studies~\citep{Lauer:2012jh,Beig:2018ga} and serves as a representation of actual applications involving liquid cavitation, such as high-pressure pumps~\citep{bohner01}. 

\begin{figure}
    \centering
    \subfigure[]{\includegraphics[height=5.5cm]{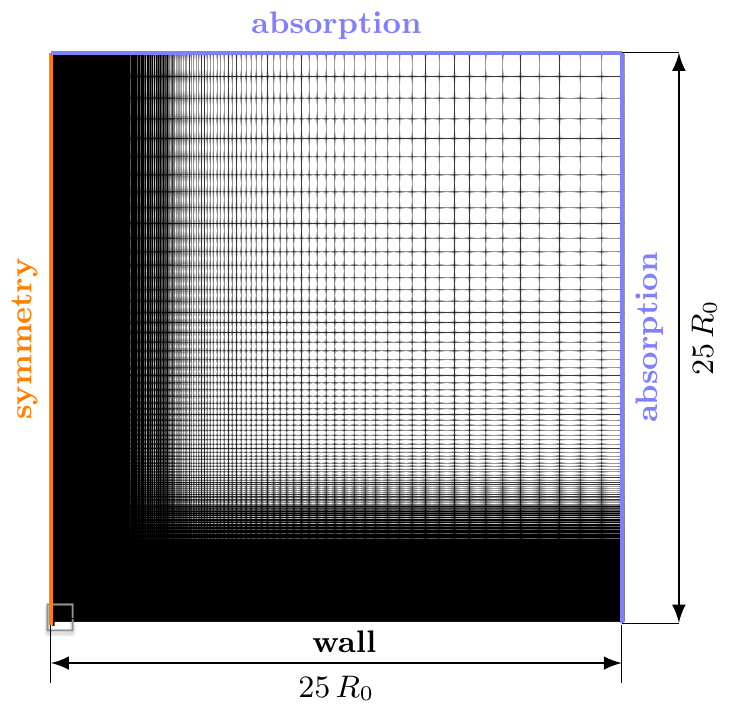}}
    \subfigure[]{\includegraphics[trim={20 0 10 0},clip,height=5.5cm]{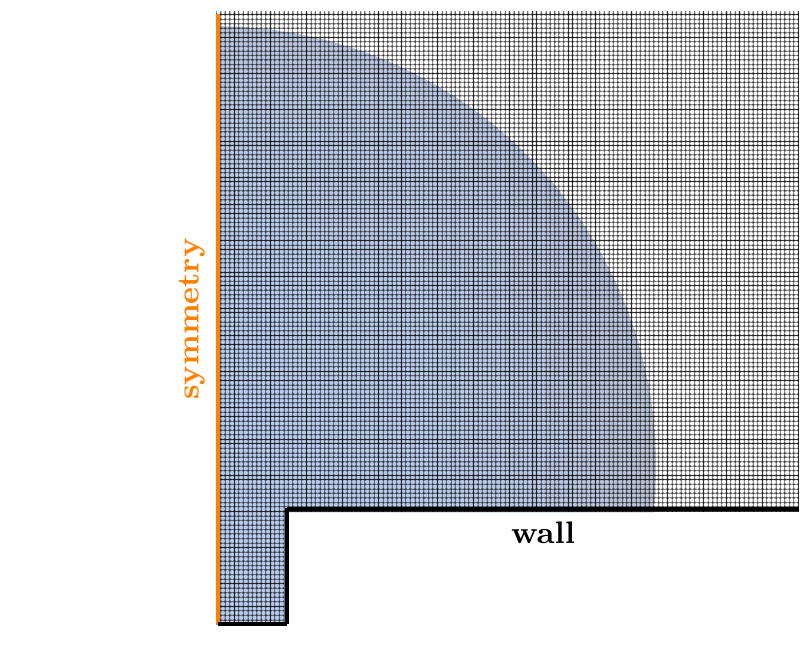}}
    \caption{
        (a) 2D axisymmetric grid configuration and boundary conditions for example case small crevice ($R_C/R_0 = 0.15$) and a stand-off distance of $S/R_0=0.35$. (b) Magnification of the near-bubble region with the bubble shaded. Only every fourth grid-line is shown in each coordinate direction.
    }
    \label{fig:grid}
\end{figure}%
Figure~\ref{fig:grid} shows the computational grid. The bubble collapse process is assumed to be axisymmetric with radial coordinate $r$, and thus a 2D axisymmetric domain of radius and length $25 R_0$ is used, matching that of previous studies of smooth-wall collapse~\citep{Lauer:2012jh}. The grid is equally spaced with 400 finite volumes per $R_0$ near the bubble (until $\hat{r} = 1.5 R_{0}$) and is progressively stretched farther from the bubble with a stretching factor of $1.01$ in each direction. This resolution has been shown to be sufficient for the conditions considered here~\citep{Lauer:2012jh,Pohl:2015keb,Beig:2018ga}. Absorbing boundary conditions are used at the outer boundaries to suppress reflecting pressure waves at these locations~\citep{toro97}. A constant CFL number of $0.4$ is used, which corresponds to a time step of $\Delta t \approx \SI{0.15}{\nano\second}$. The total simulation time is $\SI{6}{\micro\second}$, or about $1.5 t^{\ast}$
where 
\begin{gather}
    t^{\ast} = R_0 \sqrt{\frac{\rho_l }{ \Delta p}}    
    \label{e:collapse}
\end{gather}
is an estimate of the collapse time of a bubble collapse near a solid wall \citep{Plesset:1971hu}, where $\Delta p \equiv p_\infty - p_B$ is the driving pressure difference. The wall has a retarding effect on the collapse and thus $t^{\ast}$ is longer than the Rayleigh collapse time for spherical collapses ($t_\mathrm{Rayleigh}=0.915\,t^{\ast}$). Velocity and pressure are normalized as
\begin{gather}   
    u^{\ast} = \sqrt{\frac{\Delta p}{\rho_l}}, 
    \quad \text{and} \quad
    p^{\ast} = c_l \sqrt{\rho_l \Delta p},
\end{gather}
where $c_l$ is the liquid speed of sound. 

\section{Results}
\label{sec:Results}

\subsection{Considered configurations}
\label{s:cases}

\begin{figure}
    \centering
    \includegraphics[width=0.8\linewidth]{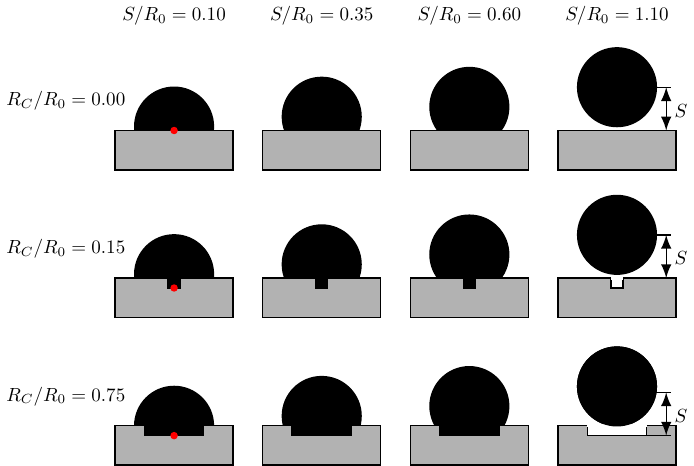}
    \caption{
        Overview of the investigated configurations. The red circle shows the $r = 0$ wall-centred position used to observe the pressure impact. Rows correspond to constant crevice size $R_C/R_0$ and columns correspond to constant stand-off distance $S/R_0$. The stand-off distance $S$ is also shown; its definition is modified to be measured from the bottom of the crevice for the $R_C/R_0 = 0.75$ cases. 
    }
    \label{fig:configs}
\end{figure}

We use stand-off distances $S/R_{0} = 0.1$, $0.35$, $0.6$, and $1.1$ (wall-detached). For each stand-off distance we consider a smooth wall ($R_C= 0$), a small crevice ($R_C/R_0= 0.15$) and a large crevice ($R_C/R_0= 0.75$), as shown in figure~\ref{fig:configs}.

We first analyse the collapse behaviour of wall-attached bubbles by increasing crevice size (smooth wall in section~\ref{subsec:rc000}, small crevice in section~\ref{subsec:rc060} and large crevice in section~\ref{subsec:rc300}), and then consider detached bubbles in section~\ref{subsec:s440}. In section~\ref{subsec:pmax}, we compare the pressure impact on the wall for all configurations and assess the cavitation erosion potential. 

\subsection{Smooth-wall-attached bubble collapse $R_C=0$ }
\label{subsec:rc000}

\begin{figure}
    \centering
    \includegraphics[width=\linewidth]{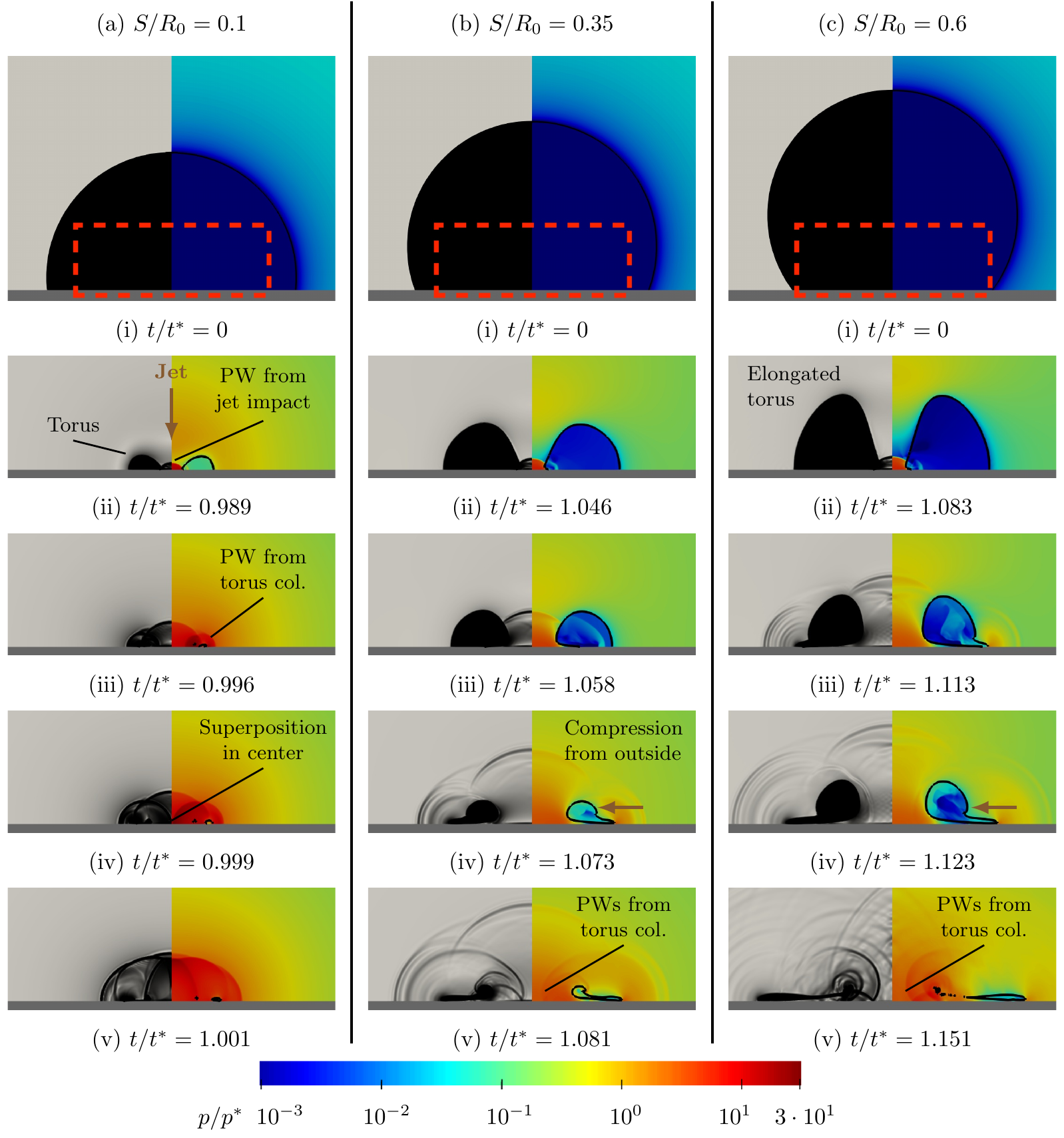}
    \caption{
        Numerical schlieren (left) and log-scale pressure fields (right) of an air bubble collapsing onto a smooth wall of varying stand-off distances $S/R_{0}$ (a)--(c) at selected times (i)--(v). Gas volume fraction $\alpha_g$ is shown as a shaded area of decreasing opacity with decreasing $\alpha_g$ (left), while the $\alpha_g = 0.5$ bubble interface is shown as a solid curve (right). (ii)--(v) are magnified to the -- -- -- rectangular region of (i). Selected pressure waves (PW) and collapse dynamics (col.) are also identified.
        }  
    \label{fig:rc000_ts}
\end{figure}

\begin{figure}
    \centering
    \includegraphics[width=\linewidth]{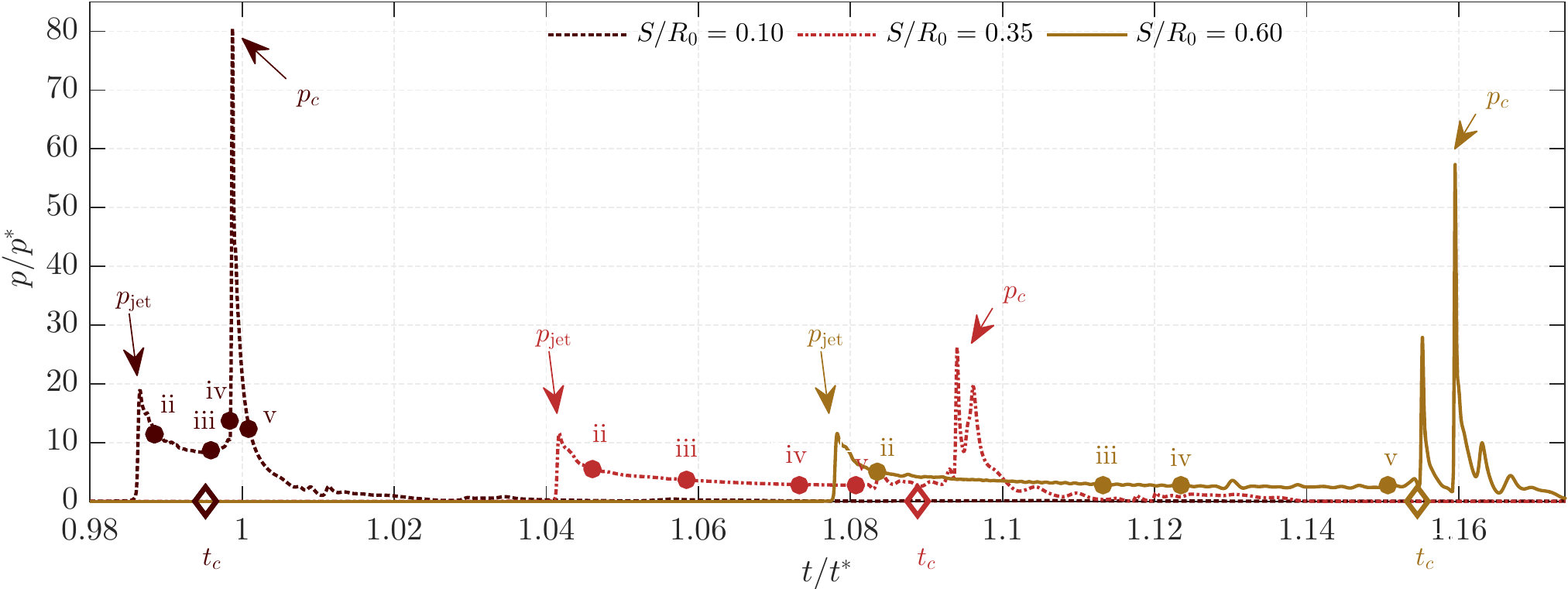}
    \caption{
        Evolution of the wall-centred pressure for the smooth-wall case at varying stand-off distances $S/R_0$. The time instances shown in figure~\ref{fig:rc000_ts} are highlighted and labelled with the corresponding row (ii--v). The pressure peaks induced by the jet impact $p_\mathrm{jet}$ and the collapse $p_c$ are indicated as such. The collapse time $t_c$ is plotted as a diamond on the $x$-axis.
    }
    \label{fig:rc000_p}
\end{figure}

Figure~\ref{fig:rc000_ts} visualizes the flow of a collapsing wall-attached bubble using the pressure field $p$ and numerical schlieren $\Phi$~\citep{Quirk:1996hv} as
\begin{gather}
    \Phi = \exp\left(- \frac{ k |\nabla \rho |}{\max |\nabla \rho|} \right),
    \label{eq:num_schlier}
\end{gather}
where $k = 400$ is used to ensure waves in the liquid are visible~\citep{Johnsen:2007wr, Meng:2018cw}. The corresponding pressures at the centre of the wall are shown in figure~\ref{fig:rc000_p}.

\begin{figure}
    \centering
    \includegraphics[width=0.99\linewidth]{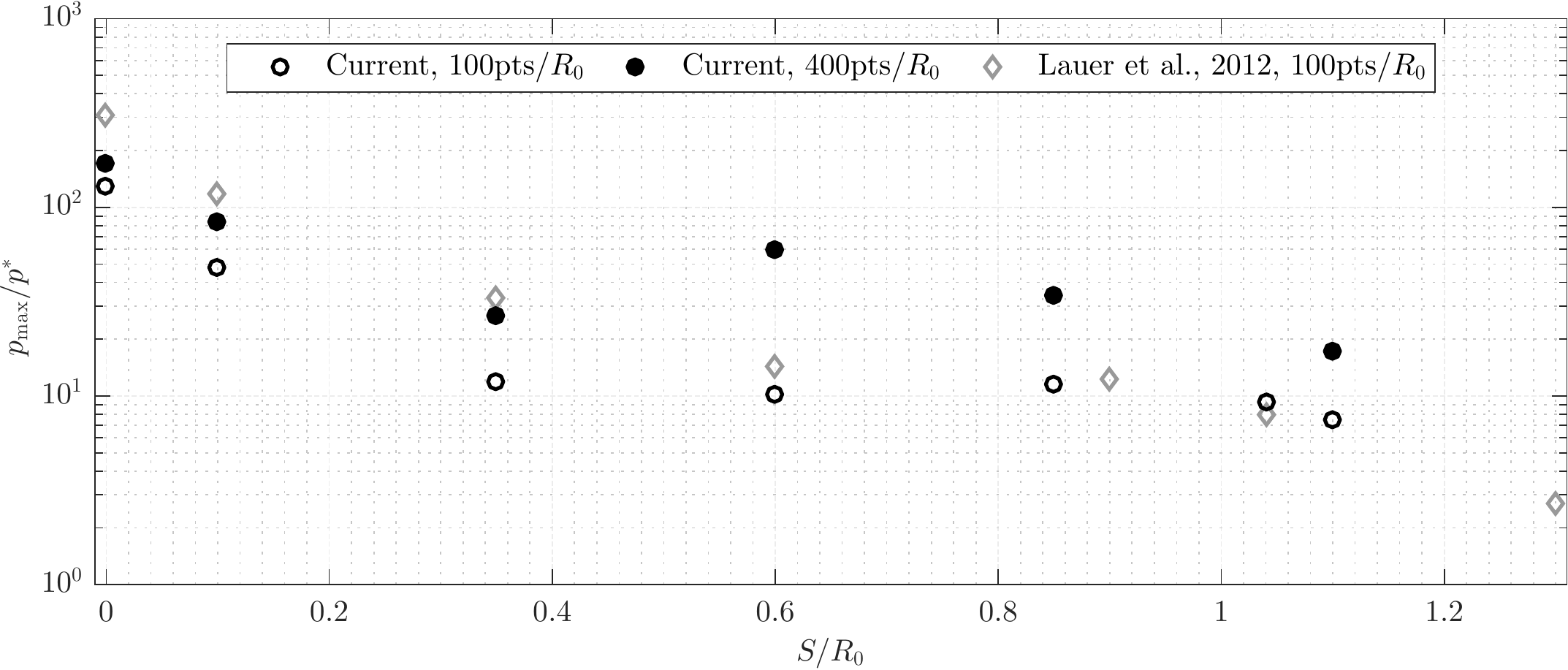}
    \caption{
        Maximum wall pressure for a smooth-wall-attached bubble of varying stand-off distance $S/R_0$ and grid resolution as labelled. Results from \citet{Lauer:2012jh} are also shown for comparison.  
    }
    \label{fig:rc000_val}
\end{figure}

For all cases a wall-directed jet is formed during the initial collapse phase. The jet impinges on the wall (row ii) leading to a pressure wave. At subsequent times the remaining toroidal bubble continues to collapse, emitting a pulse that travels radially inward and collides at $r = 0$. 

The collapse of the torus becomes increasingly non-uniform, with a portion near the wall being pinched away from the main torus. Pressure waves emitted near the pinching location are evident, starting in (b,iv) and (c,iii) respectively. In addition, a compression of the torus from the outside pushes its upper part towards the centre (b,iv), (c,iv). During the final collapse phase, two pressure waves propagate inward, focus, and result in two distinct pressure pulses at the wall centre, as visible in figure~\ref{fig:rc000_p}.

The impact of a liquid jet onto the wall generates a water hammer pressure proportional to the jet velocity $p_\mathrm{jet}\propto \rho_l\,c_l\,u_\mathrm{jet}$. The jet-induced pressure peak $p_\mathrm{jet}$ is clearly visible from the wall-centred pressure signals of figure~\ref{fig:rc000_p}. For $S/R_0 =0.1$, the peak is about twice as high as for the others. The high jet velocity at this small stand-off distance is a result of the bubble shape being almost hemispherical. A hemispherical bubble attached to an inviscid wall collapses like a spherical bubble with a uniform and high acceleration of the interface. For $S/R_0 =0.1$ the initial stages of the collapse resemble those of a collapsing spherical bubble, with the formation of the liquid jet immediately preceding the total collapse and the jet reaching a high velocity. Similar observations were made by \citet{Philipp:1998eg}, who also experimentally recorded the highest jet-induced pressures at small stand-off distances.

In the configurations considered, the total collapse is the collapse of the gas torus. We determine the collapse time $t_c$ by the minimum gas volume. The pressure waves emitted at total collapse result in collapse-induced pressure peaks $p_c$ (see figure~\ref{fig:rc000_p}). Thus, the jet impact on the wall as well as the shock waves emitted during total collapse cause high pressure peaks and potentially material damage. For the rough wall cases, we also observe pressure peaks induced by post-collapse wave dynamics. In section~\ref{subsec:pmax}, we compare these three pressure peaks for all configurations. For the smooth-wall cases, $p_c$ is significantly higher than $p_\mathrm{jet}$, which agrees with the findings of \citet{Lauer:2012jh}.

In figure~\ref{fig:rc000_val} the maximum wall pressure $p_\mathrm{max}$ is compared with that of~\citet{Lauer:2012jh} for our present resolution ($400 \text{pts}/R_0$) and $100 \text{pts}/R_0$, which matches their study. The current results follow the same trends, although with lower pressures for the attached-bubble cases ($S < R_0$). The maximum pressure is known to be sensitive to resolution, although a discrepancy also exists for identical grid resolutions ($100 \text{pts}/R_0$). \citet{Lauer:2012jh} consider condensation, while we model the bubble content as non-condensable gas. The damping of the maximum pressure observed is consistent with previous analysis of bubbles containing non-condensable gas~\citep{Trummler:2018ww, Pishchalnikov:2018pp}.

\subsection{Small crevice $R_C/R_0 = 0.15$}
\label{subsec:rc060}

\begin{figure}
    \centering
    \includegraphics[width=\linewidth]{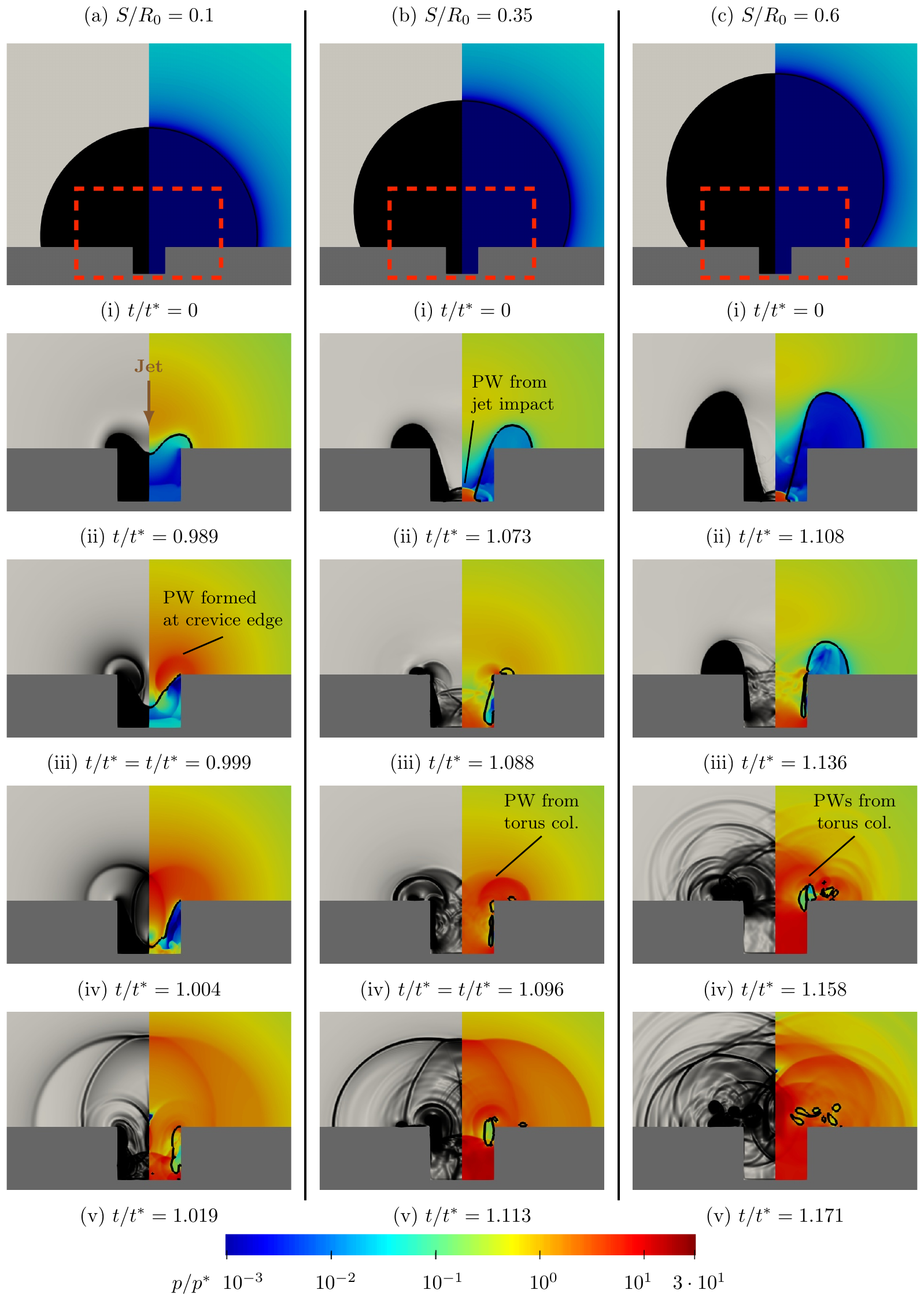}
    \caption{
        Numerical schlieren (left) and pressure fields (right) of an air bubble collapsing onto a wall with a small crevice $R_C/R_{0}=0.15$ at varying stand-off distances $S/R_{0}$ (a)--(c) at selected times (i)--(v). (ii)--(v) are magnified to the -- -- -- rectangular region shown in (i). Selected pressure waves (PW) and collapse dynamics (col.) are also identified. 
        }
    \label{fig:rc060_ts}
\end{figure}

\begin{figure}
    \centering
    \includegraphics[width=\linewidth]{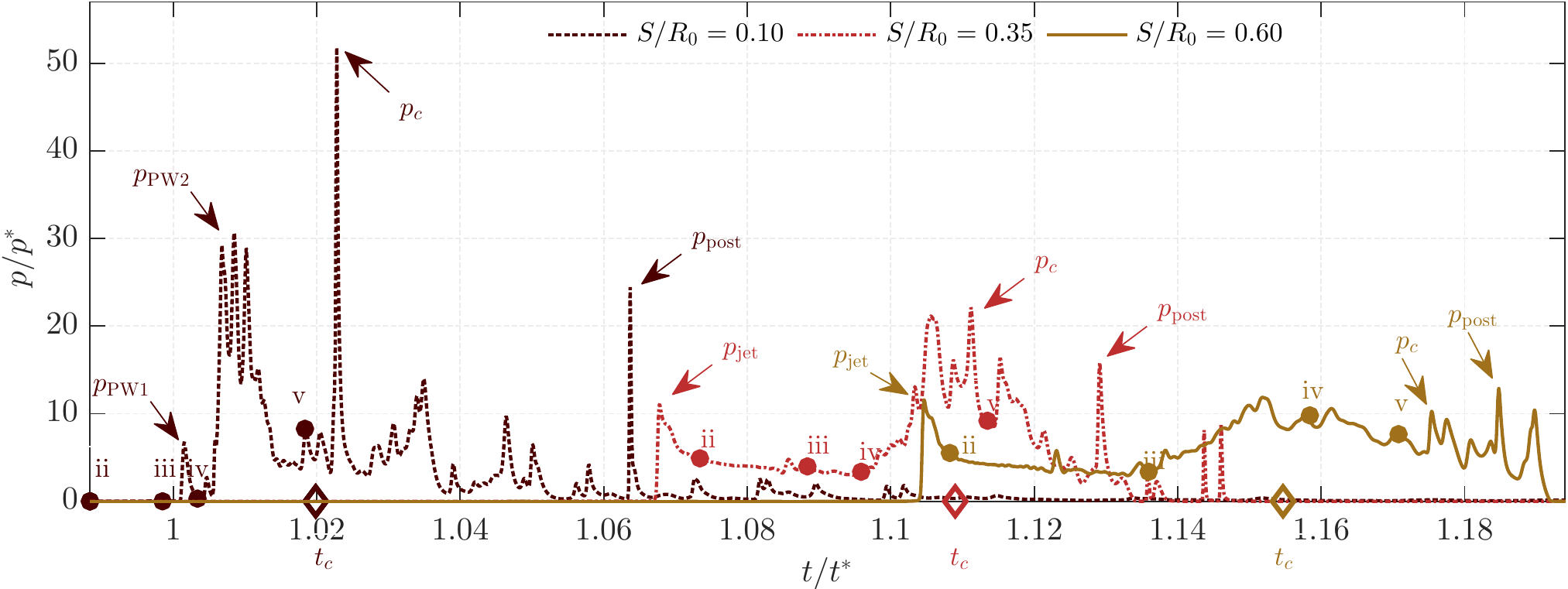}
    \caption{
        Evolution of the wall-pressure at $r = 0$ for the case $R_C/R_0 = 0.15$ at varying stand-off distances $S/R_0$. The time instances shown in figure~\ref{fig:rc060_ts} are highlighted and labelled with the corresponding row (ii--v). The pressure peaks induced by the jet impact $p_\mathrm{jet}$, the pressure-wave $p_\mathrm{PW}$, the collapse $p_c$ and post-collapse wave dynamics $p_\mathrm{post}$ are indicated as such. The collapse time $t_c$ is plotted as a diamond on the $x$-axis. 
    }
    \label{fig:rc060_p}
\end{figure}

Visualizations of a collapsing $R_C/R_0 = 0.15$ crevice-wall-attached bubble at varying stand-off distances $S/R_0$ are shown in figure~\ref{fig:rc060_ts} and the corresponding wall pressures are shown in figure~\ref{fig:rc060_p}. 

For the smallest stand-off distance case ($S/R_0 = 0.1$), the initial stages of the collapse match that of the smooth-wall cases, with a jet piercing the bubble and generating a toroidal structure. However, in this case the gas torus is ultimately fully contained in the crevice. As shown in (a,iii), a pressure wave is emitted when the liquid has reached the sharp edge of the crevice and is suddenly stopped there. This wave propagates radially outwards (a,iv) and collides in the centre inducing a small pressure peak at the wall-centre, see $p_\mathrm{PW1}$ in figure~\ref{fig:rc060_p}. The pressure wave continues to travel towards the other crevice side pushing the gas away from the crevice bottom and pressing it against the opposite side wall (a,v). Between (a,iv) and (a,v) the pressure wave and its reflections induce high pressure fluctuations at the wall centre ($p_\mathrm{PW2}$). The last time step depicted (a,v) is close to the final collapse, which causes the highest pressure peak. 

For the larger stand-off distances $S/R_0 = 0.35$ and $0.6$, the jet penetrates the entire bubble and hits the crevice bottom. A gas torus remains on the upper wall and a gas layer covers the side walls. Like in the smooth wall cases, the gas torus outside of the crevice collapses ((b,iv), (c,iv)), emitting intense pressure waves. These waves propagate radially outward, interfere with each other, and are reflected within the crevice. The time steps (b,v) and (c,v) both visualize the complex wave pattern after the total collapse. 

Figure~\ref{fig:rc060_p} shows that the wall-centred pressures associated with the $S/R_0 = 0.35$ and $0.6$ cases are qualitatively similar. Both have a pressure peak due to the jet impact, followed by a time-delayed accumulation of pressure peaks during and after the final collapse phase. For $S/R_0 = 0.6$ these pressure peaks are smaller since the intense pressure waves are more concentrated in the area above the crevice (see (c,iv,v)) and thus decay until they reach the crevice bottom. 

At all stand-off distances, significant pressure peaks are induced by post-collapse wave dynamics (see $p_\mathrm{post}$ in figure~\ref{fig:rc060_p}). 

\subsection{Large crevice $R_C/R_0 = 0.75$}
\label{subsec:rc300}

\begin{figure}
    \centering
    \includegraphics[width=0.85\linewidth]{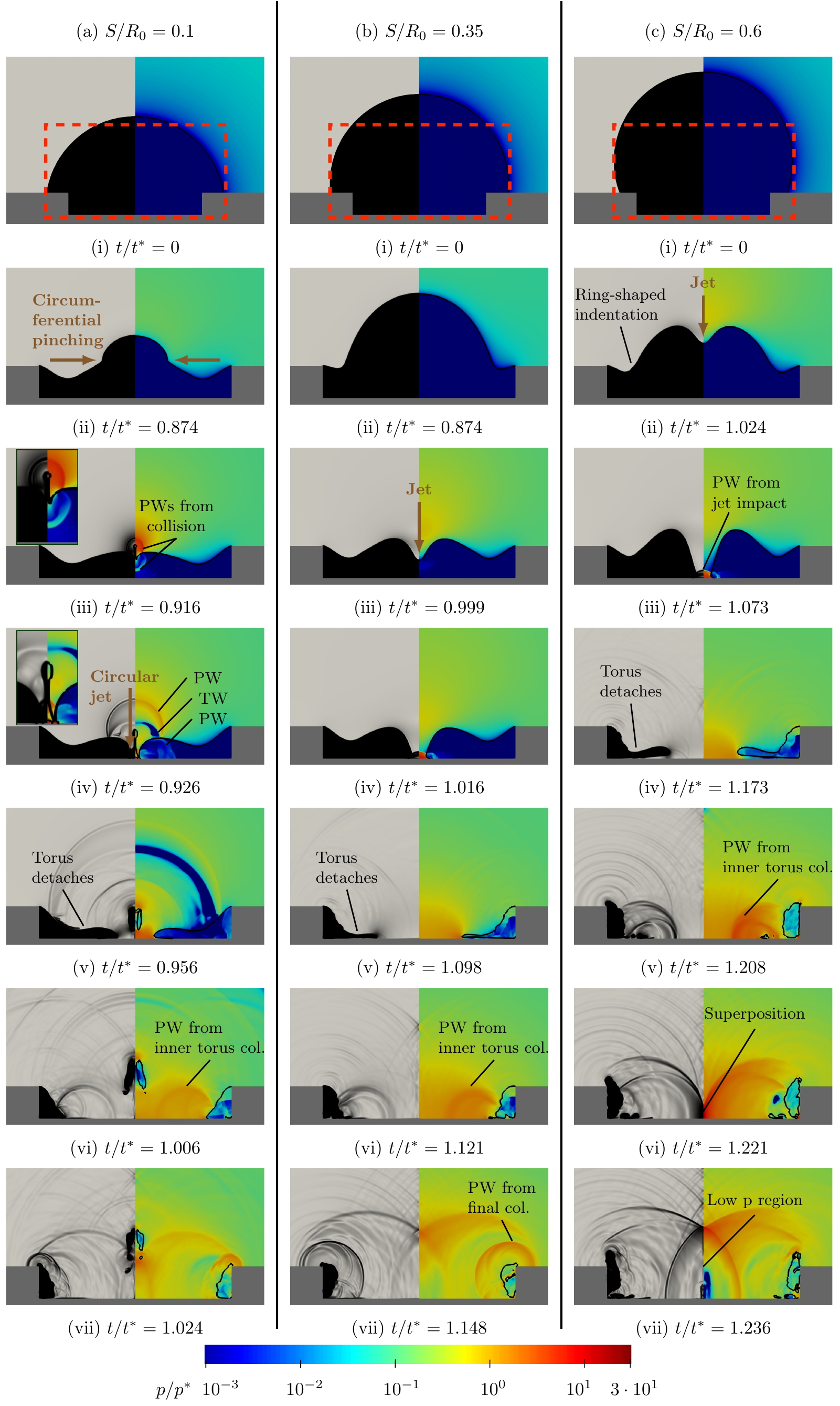}
    \caption{
        Numerical schlieren (left) and pressure fields (right) of an air bubble collapsing onto a wall with crevice size $R_C/R_{0}=0.75$ at varying stand-off distances $S/R_{0}$ (a)--(c) at selected times (i)--(vii). In (a,iii) and (a,iv) the relevant areas are additionally magnified in the upper left corner. 
        (ii)--(vii) are magnified to the -- -- -- rectangular region shown in (i).
        Selected pressure waves (PW), tension waves (TW), and collapse dynamics (col.) are also identified.
     }
    \label{fig:rc300_ts}
\end{figure}

\begin{figure}
    \centering
    \includegraphics[width=\linewidth]{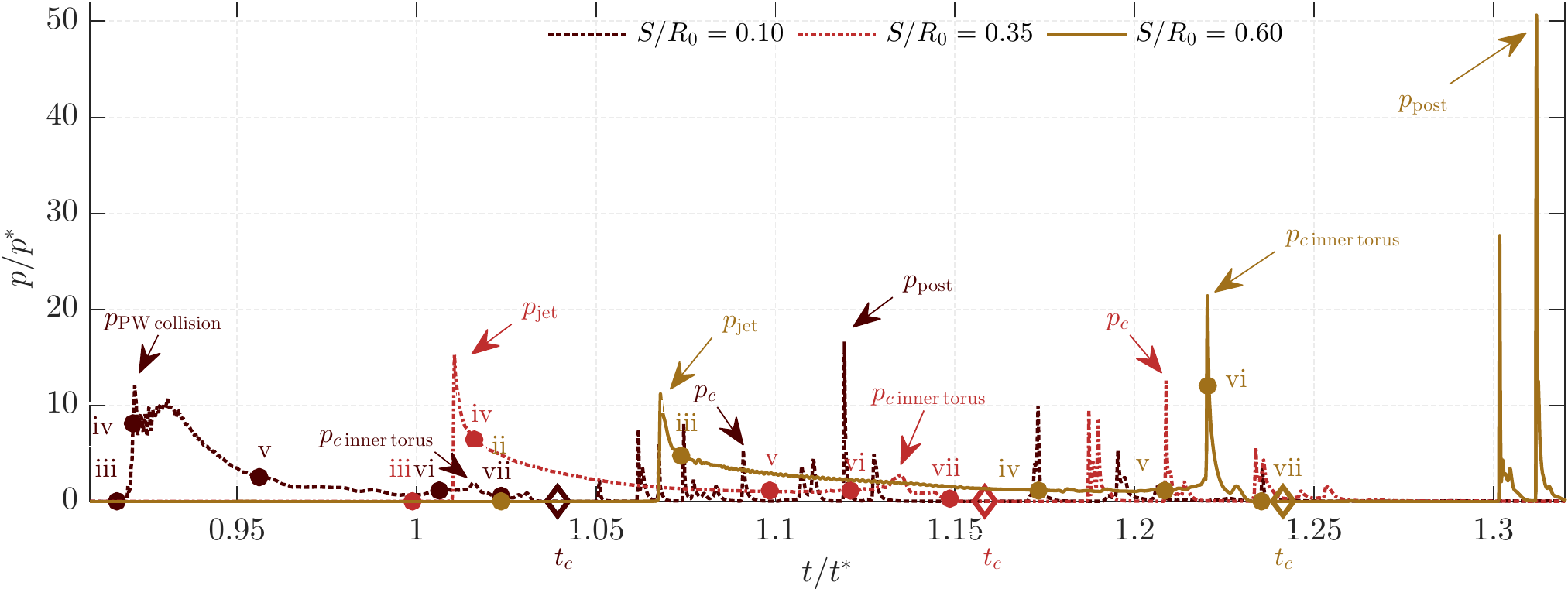}
    \caption{
        Evolution of the wall-pressure at $r = 0$ for the case $R_C/R_0 = 0.75$ at varying stand-off distances $S/R_0$.
        The time instances shown in figure~\ref{fig:rc300_ts} are highlighted and labelled with the corresponding row (ii--vii). The pressure peaks induced by the jet impact $p_\mathrm{jet}$, the collision of the pressure-wave $p_\mathrm{PW\, collision}$, the collapse of the inner torus $p_{c\,\mathrm{inner\,torus}}$, the total collapse $p_c$ and post-collapse wave dynamics $p_\mathrm{post}$ are indicated as such. The collapse time $t_c$ is plotted as a diamond on the $x$-axis. 
    }
    \label{fig:rc300_p}
\end{figure}

We next consider the large crevice $R_C/R_0 = 0.75$ cases. Recall that $S$ is now measured from the bottom of the crevice wall to the bubble centre, instead of from the top of the crevice wall. Figure~\ref{fig:rc300_ts} visualizes the collapses and the corresponding wall-centred pressures are shown in figure~\ref{fig:rc300_p}. 

For the $S/R_0 = 0.1$ case (column a), the fraction of the bubble surface initially exposed to the high-pressure liquid is comparable to that of a bubble with a small negative stand-off distance ($S/R_0 - d/R_0 = 0.1 - 0.25 = - 0.15$). Consequently, the initial collapse phase resembles that of such a configuration. \citet{Lauer:2012jh} and \citet{Shima:1977df} report a collapse behaviour similar to that of a spherical collapse with an additional circumferential pinching at the position of maximum extension, resulting in a mushroom-shape. Here, (a,ii) shows the compressed upper part of the bubble and also a circumferential pinching. Additionally, a ring-shaped indentation of the bubble can be observed. 

The circumferential pinching meets at the $r = 0$ axis of symmetry, generating a pressure wave (a,iii), which propagates radially outward in the liquid and the gas. When the pressure wave in the gas reaches the bottom wall, it induces a pressure peak there (see figure~\ref{fig:rc300_p}, $p_\mathrm{PW\,collision}$). The pressure wave in the liquid is partially reflected at the gas-liquid interface, and generates a tension wave following the initial pressure wave (a,iv,v). Furthermore, the collision of the circumferential pinching results in the formation of a wall-normal circular jet, see (a,iii,iv). The subsequent circular jet impacts on the bottom wall and pushes away the gas in the crevice centre. A secondary bubble pinches off and moves upwards (a,v). From the remaining flattened gas torus, an inner gas torus detaches at the position of the ring-shaped indentation, collapses (a,vi) and emits a pressure wave propagating in the direction of $r = 0$ (a,vii). At the same time, the remaining gas is pressed towards the crevice side walls and pressure waves are formed at the sharp edges of the crevice (a,vii). 

For the $S/R_0 = 0.35$ and $0.6$ cases (figure~\ref{fig:rc300_ts}~(b) and (c)), a ring-shaped indentation forms close to the crevice edge during the initial collapse phase, similar to that of the $S/R_0 = 0.1$ case.  In addition, the jet indents the bubble from the top, as observed for the small crevice and the smooth wall configurations. (b,iv) and (c,iii) show that the larger stand-off distance results in a more curved bubble interface when the jet impacts the wall. Similar to the $S/R_0 = 0.1$ case, an inner torus detaches from the main torus at the position of the ring-shaped indentation ((b,v) and (c,iv)) and collapses, emitting a pressure wave ((b,vi) and (c,v)). The pressure wave propagates to the centre, collides there inducing a pressure peak (c,vi) ($p_{c\,\mathrm{inner\, torus}}$) and then continues, resulting in a low-pressure area (c,vii). This pressure decrease can cause a vapour bubble rebound when phase-change processes are taken into account. The final collapse occurs when the remaining gas torus in the corner of the crevice is compressed to its minimum size (b,vii).

The pressure signals in figure~\ref{fig:rc300_p} show the jet-induced pressure peak $p_\mathrm{jet}$ for $S/R_0=0.35$ and $S/R_0=0.6$. For $S/R_0=0.35$ $p_\mathrm{jet}$ is higher because the initially liquid-exposed part of the bubble interface is almost a hemisphere and is thus strongly accelerated, see also section~\ref{subsec:rc000}. For $S/R_0=0.1$, there is no jet-induced pressure peak in the centre due to the circular jet. However, a pressure peak of about the same intensity is induced by the pressure wave emitted when the circumferential pinching collides ($p_\mathrm{PW\,collision}$).

This first peak is followed by a peak $p_{c\,\mathrm{inner\,torus}}$ caused by the collapse of the inner detached torus. As $S/R_0$ increases, this pressure peak increases since the volume of the detached inner torus increases, resulting in a stronger pressure wave. Due to the preceding collapse of the inner torus, a smaller gas volume is associated with the final collapse phase. Furthermore, the collapse occurs at the crevice corner, and thus the induced pressure waves are less intense at the wall-centre. As a result, the collapse-induced pressure peak in the centre $p_c$ is comparatively small and is exceeded by $p_\mathrm{jet}$ (or respectively by $p_\mathrm{PW\,collision}$). Indeed, for $S/R_0=0.6$, the total collapse does not generate a pressure peak at the wall centre. 

After the final collapse, intense wave dynamics occur, which can lead to high pressure peaks. For $S/R_0 = 0.1$ and $0.6$, these post-collapse pressure peaks $p_\mathrm{post}$ are the maximum pressure observed.

\subsection{Collapse of a wall-detached bubble ($S/R_0 = 1.1$)}
\label{subsec:s440}

\begin{figure}
    \centering
    \includegraphics[width=\linewidth]{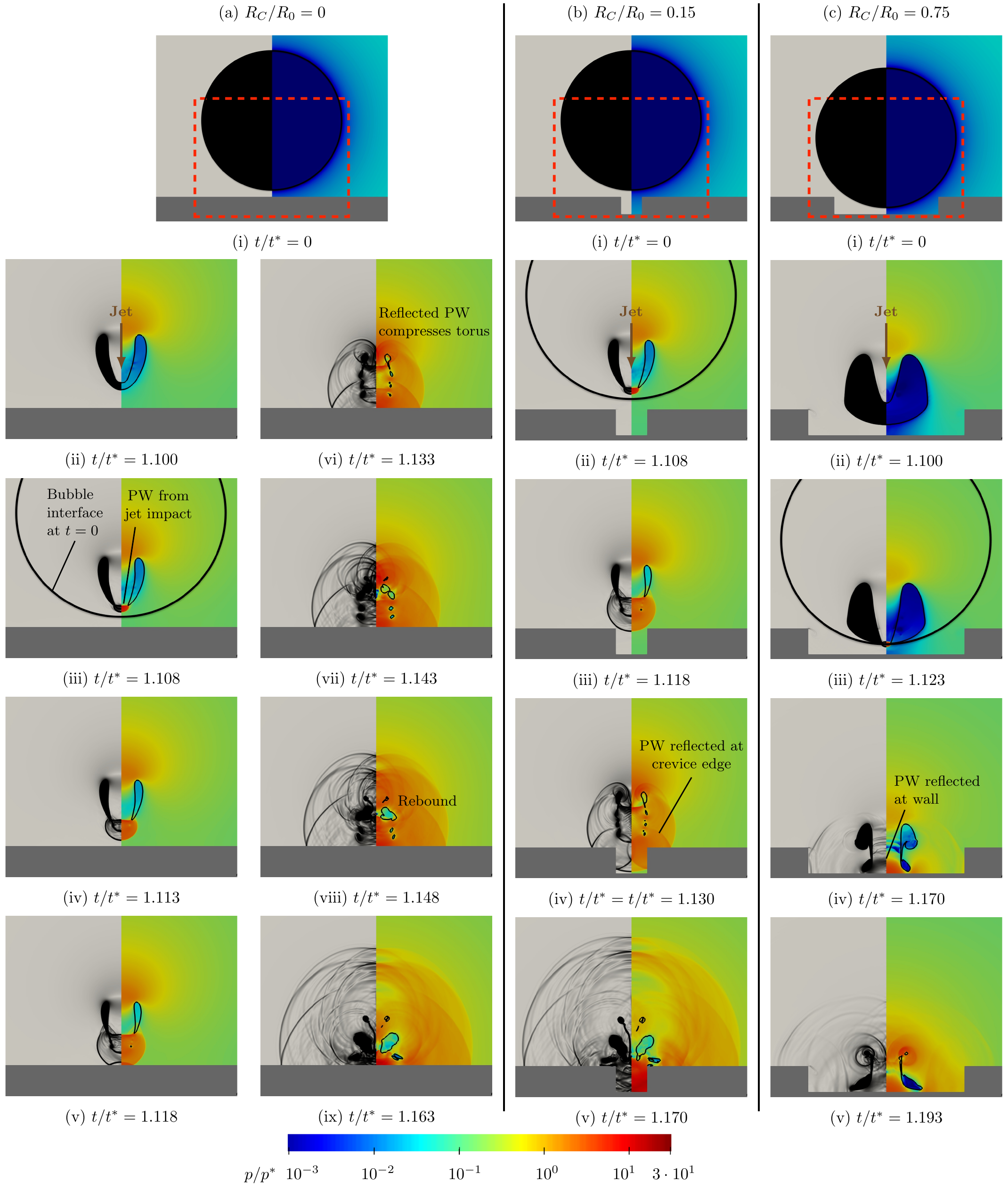}
    \caption{
        Numerical schlieren (left) and pressure fields (right) of a wall-detached air bubble ($S/R_0 = 1.1$) collapsing onto a wall of varying crevice size $R_C/R_{0}$ (a)--(c) at selected times as labelled.
        (ii)--(ix) are magnified to the -- -- -- rectangular region shown in (i). The solid curve in (a,iii), (b,ii) and (c,iii) indicates the initial position of the bubble interface. 
        Selected pressure waves (PW) are also identified. 
        }  
    \label{fig:s440_ts}
\end{figure}

\begin{figure}
    \centering
    \includegraphics[width=\linewidth]{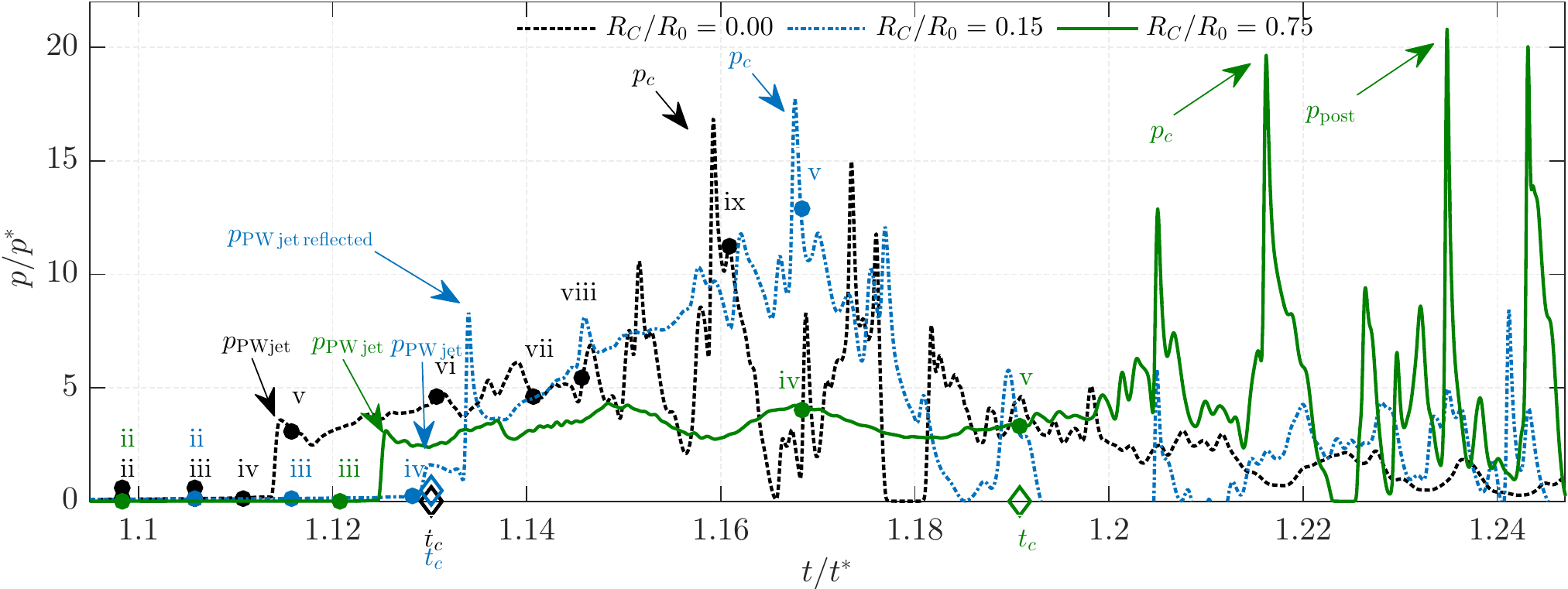}
    \caption{
        Evolution of the wall-pressure at $r = 0$ for the $S/R_0 = 1.1$ case and varying crevice sizes $R_C/R_0$. The time instances shown in figure~\ref{fig:rc300_ts} are highlighted and labelled with the corresponding row (ii--v). The pressure peaks induced the pressure-wave $p_\mathrm{PW\, jet}$, the collapse $p_c$ and post-collapse wave dynamics $p_\mathrm{post}$ are indicated as such. The collapse time $t_c$ is plotted as a diamond on the $x$-axis. 
    }
    \label{fig:s440_p}
\end{figure}

The collapse of wall-detached bubbles ($S/R_0 = 1.1$) are visualized in figure~\ref{fig:s440_ts} for varying crevice sizes. The corresponding wall-centred pressure evolution is shown in figure~\ref{fig:s440_p}. As observed for previous cases, the aspherical pressure distribution leads to an indentation of the top of the bubble and the formation of a jet penetrating the bubble for all cases. The monitored jet velocities are about $u_\mathrm{jet}/u^{\ast} \approx 10$, which is in good agreement with previous studies for smooth walls~\citep{Lauer:2012jh,Supponen:2016jnb}.

For the smooth-wall case (figure~\ref{fig:s440_ts}~(a)) the jet impacts the far-side bubble interface at $t = 1.1t^{\ast}$ and a pressure wave is emitted (a,iii). The impact time of the jet at the bubble wall and the bubble position with respect to the initial configuration are in good agreement with previous observations~\citep{Supponen:2016jnb}. The jet impact results in an upward and a downward moving wave front (see (a,iv)), with the latter being curved. The numerical schlieren shows an additional downward moving density jump corresponding to a contact wave. When the downward moving pressure wave impacts the wall, a pressure peak is induced (see also figure~\ref{fig:s440_p}, $p_\mathrm{PW\,jet}$). The pressure wave is then reflected at the wall (a,v), compressing the remaining bubble torus from bottom to top (a,vi) leading to the total collapse. After the collapse (a,vii--ix), a gas torus rebounds and moves towards the wall. 

Figure~\ref{fig:s440_ts}~(b) shows that the small crevice does not significantly change the collapse and rebound behaviour compared to the smooth wall. The main difference is the reflection of the pressure wave emitted at jet-bubble-impact at the crevice edge (b,iii--iv) and the resulting different wave patterns. 

For $R_C/R_0 = 0.75$ (figure~\ref{fig:s440_ts}~(c)), the crevice initially suppresses the compression of the lower part of the bubble, resulting in a different shape during jet penetration, at jet impact, and also after compression by the reflected wave (c,ii--v). Furthermore, this increases the collapse time by about 5\% when compared to the smooth-wall case.

The pressure signals (figure~\ref{fig:s440_p}) show that the pressure wave due to the jet-bubble-impact results in a pressure peak $p_\mathrm{PW\,jet}$ for all configurations. For the small crevice, the pressure wave has to pass a longer distance and thus the peak is smaller. However, the reflection and superposition of the wave at the edge of the crevice results in a more intense peak following ($p_\mathrm{PW\,jet\,reflected}$). 

After the collapse, all three pressure signals exhibit pressure fluctuations with significant peaks that exceed $p_\mathrm{PW\,jet}$. For the large crevice, these peaks are modestly higher than those of the other cases, since the collapse, the rebound and the associated wave dynamics take place closer to the wall. In addition, there are pressure peaks induced by post-collapse wave dynamics for the large crevice. 

\subsection{Assessment of cavitation erosion potential}
\label{subsec:pmax}

\begin{figure}
    \centering
    \includegraphics[width=\linewidth]{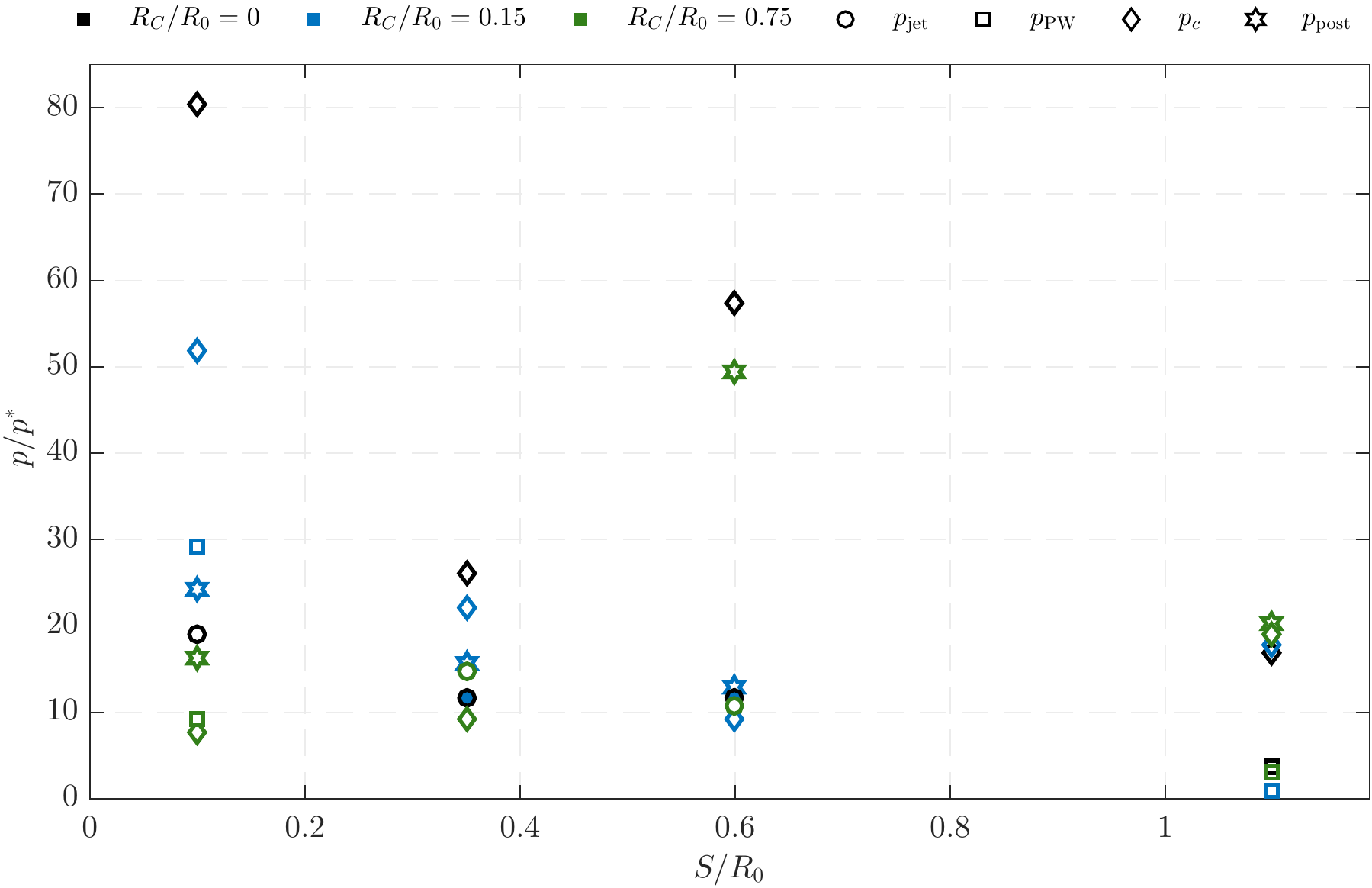}
    \caption{
        Pressure peaks observed at the wall centre ($r = 0$) over the stand-off distance $S/R_0$ for varying crevice sizes $R_C/R_0$, indicated by different colours. The jet-induced pressure $p_\mathrm{jet}$, collapse-induced pressure $p_c$, and pressure from the subsequent wave dynamics $p_\mathrm{post}$ are shown. For cases with no $p_\mathrm{jet}$, the pressure peak induced by initial pressure waves $p_\mathrm{PW}$ is shown instead. Note that no $p_\mathrm{post}$ is observed for smooth wall configurations. 
    }
    \label{fig:pressure_impacts_rc_}
\end{figure}

\begin{figure}
    \centering
    \includegraphics[width=\linewidth]{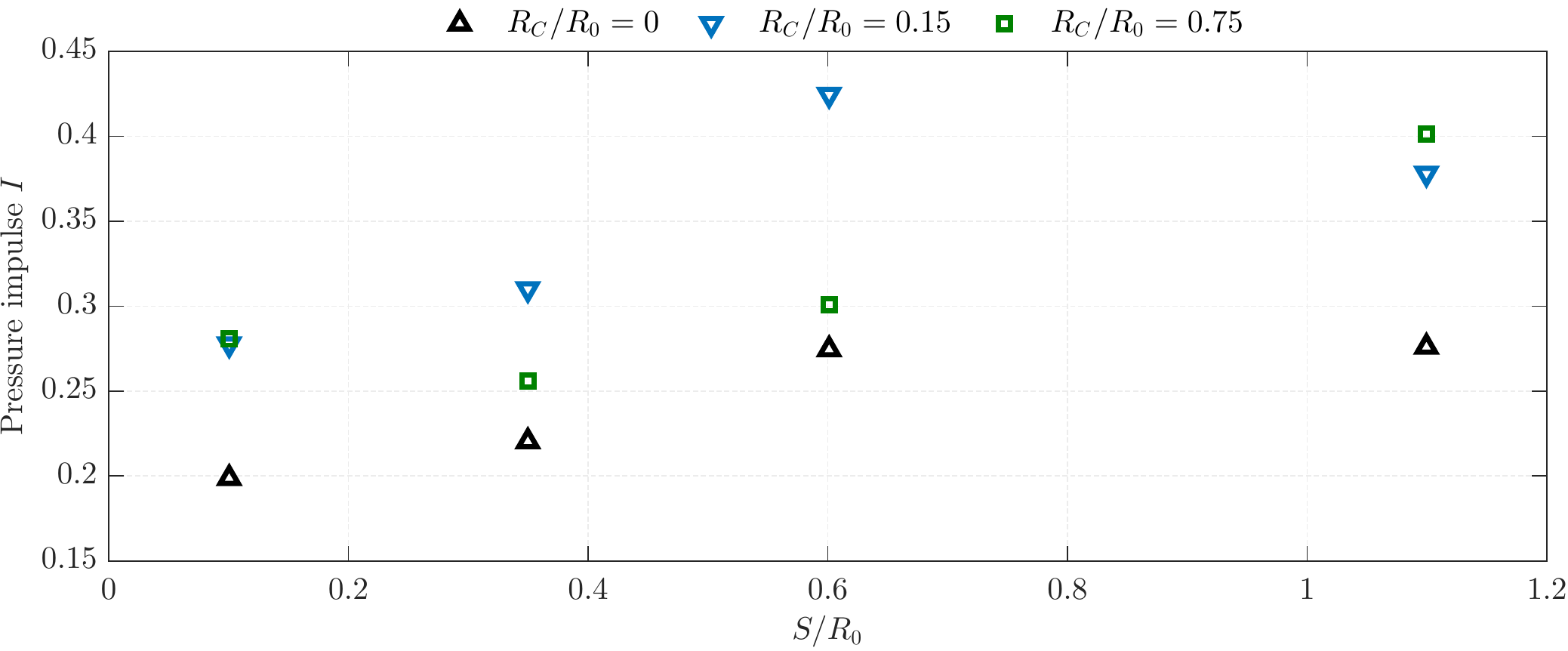}
    \caption{
        Pressure impulse $I$ at the wall centre ($r = 0$) over the stand-off distance $S/R_0$ for varying crevice sizes $R_C/R_0$, indicated by different colours and symbols. 
        }  
    \label{fig:momentum}
\end{figure}

\begin{figure}
    \centering
    \includegraphics[width=\linewidth]{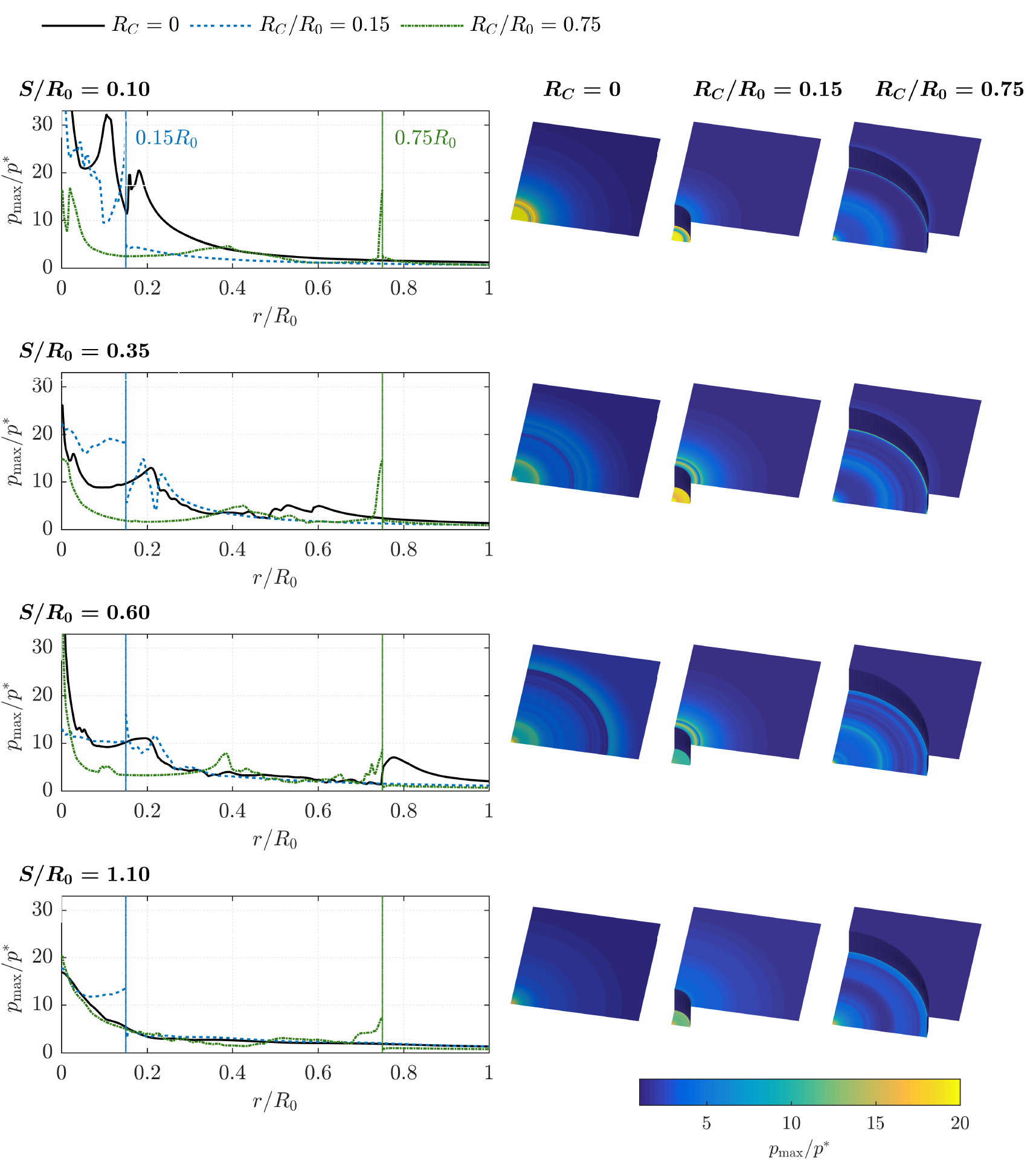}
    \caption{
        Maximum wall pressure $p_\mathrm{max}$ of the entire bubble-collapse process
        for varying radial locations $r$ with rows corresponding to the stand-off distances $S/R_0$. 
        First column: $p_\mathrm{max}$ over $r$, where the pressure axes are truncated to promote visibility; the maximum values over all $r$ are shown in~figure~\ref{fig:pressure_impacts_rc_}. Second to fourth column: 3$-$D visualization of the maximum wall pressure for each crevice size.
        }  
    \label{fig:pmax_plot_all}
\end{figure}

The previous sections showed that jet impact, collapse and, in certain configurations, post-collapse wave dynamics induce high pressure peaks in the crevice centre. Peak pressures are in the range of $15 p^\ast$--$80 p^\ast$, which corresponds to about $\SI{2}{\giga\pascal}$--$\SI{12}{\giga\pascal}$ for a bubble exposed to a driving pressure of $p_\infty=\SI{e7}{\pascal}$. These values significantly exceed the strengths of many common engineering materials, such as the $\SI{0.55}{\giga\pascal}$ ultimate tensile strength of stainless steel. Thus, there is potential for significant material erosion. To investigate this, we compare the pressures associated with the various collapse mechanisms, evaluate the induced pressure impulse, and analyse the spatial distribution of maximum wall pressures. 

Figure~\ref{fig:pressure_impacts_rc_} compares the wall-centred pressures associated with the various processes. The jet-induced pressure peaks $p_\mathrm{jet}$ do not vary significantly for the three wall configurations, with nearly identical values for $R_C = 0$ and $R_C/R_0 = 0.15$. At $S/R_0 = 0.1$, $R_C=0$ and at $S/R_0 = 0.35$, $R_C/R_0=0.75$, high interface accelerations and jet velocities occur, resulting in an increased $p_\mathrm{jet}$ as discussed in sections~\ref{subsec:rc000} and~\ref{subsec:rc060}. For $S/R_0 = 0.1$ with $R_C>0$ and for $S/R_0=1.1$, there are no jet-induced pressure peaks. 

The collapse-induced pressure $p_c$ is for the smooth wall higher than for the creviced configurations. At the smooth wall, the final collapse position is closer to the wall-centre ($r=0$) and a larger gas volume is associated with the final collapse phase. For the large crevice, $p_c$ is significantly smaller than that of the other configurations. In these cases, a smaller gas volume is associated with the final collapse due to a preceding collapse of an inner detached torus, see section~\ref{subsec:rc300}. Furthermore, the final collapse is in the crevice corner and thus the intensity of the pressure waves decreases until they reach $r=0$.

For the creviced configurations, high pressure peaks can be caused by the wave dynamics present after collapse. These peaks can be close to the maximum pressure induced in the smooth wall configuration (see $S/R_0 = 0.6$), indicating erosion potential. For the detached configuration all pressure impacts are of comparable intensity. 

Figure~\ref{fig:momentum} shows the pressure impulse at the crevice centre,
\begin{gather}
    I = \frac{1}{1.5\,t^{\ast}\,p^{\ast}}\int_{0}^{1.5\,t^{\ast}}\bigl(p(t)-p(t=0)\bigr) \,\mathrm{d} t.
\end{gather}
$I$ takes into account whether an increased pressure is present over a longer period of time. In contrast to the maximum wall pressure, the impulse is not biased by single instantaneous peak values. Despite the smaller maximum $p$ for the creviced cases, the impulse for these configurations is larger than that for the smooth-wall cases. For the small crevice, $I$ is about 50\% larger than at the smooth wall at all stand-off distances. 

Figure~\ref{fig:pmax_plot_all} shows the maximum wall pressure $p_\mathrm{max}$ at varying radial locations and a visualization of the $p_\mathrm{max}$ distribution. First the attached configurations are discussed by crevice size and then the detached ones. 

For the smooth wall configurations, there is a collapse-induced peak in $p_\mathrm{max}$ at the centre with a significant radial decay. In addition, modest pressure peaks are observed at about $r\approx 0.2 R_0$, where the torus collapses. This pressure distribution is in agreement with predicted damage patterns by~\citet{Philipp:1998eg}, who found ring-shaped damage ($r\approx 0.3 R_0$) and a smooth indentation at the wall centre.

For the small crevice, significant pressure peaks are induced over the entire crevice bottom. They are especially high at $S/R_0 = 0.35$, where they exceed that of the smooth wall. On the upper wall are peaks at about $r \approx 0.2 R_0$ which are related to the torus collapsing at this position (see figure~\ref{fig:rc060_ts}). For the small stand-off distance $S/R_0 = 0.1$ no increased maximum pressures are observed at the upper wall, because the collapse takes place within the crevice.

For the large crevice, the collapse of the detached gas torus results in a modest pressure peak at $r \approx 0.4 R_0$, as described in section~\ref{subsec:rc300}. This gas torus is largest for the $S/R_0 = 0.6$ case, and thus leads to highest pressures at this position. The total collapse is in the crevice corner ($r = R_C$) and induces large pressures at this location. Furthermore, at $S/R_0 = 0.1$ two pressure peaks are observed near $r = 0$. The impact of the circular jet results in the off-centre peak, while the shock wave after the collapse results in the $r = 0$ maximum pressure. 

For all detached-bubble cases, the maximum $p_\mathrm{max}$ occurs at $r = 0$, and decays with increasing $r$ apart from a modest increase at $r = R_C$. For the small crevice, there is again a high pressure impact over the entire $r< R_C$ area. Nevertheless, overall the effect of $R_C$ on $p_\mathrm{max}$ appears to decrease with increasing $S$. 

Three distinct processes can cause high pressures at the crevice walls and thus, potential damage: the jet impact, the primary collapse, and post-collapse wave interactions. For smooth-wall cases, the pressure peaks are most significant at the wall centre and cavitation erosion can be expected at this location. For the small crevice cases, a high pressure occurs across the entire crevice bottom, leading to a broader area of possible cavitation erosion. For the large crevice cases, the pressure peaks seen at the crevice corners are also significant, and cavitation erosion is possible at these locations as well.

\section{Conclusion}
\label{sec:Conclusion}

The collapse of a single gas bubble attached or near a smooth or creviced surface was investigated using high-resolution simulations.
Variations of the stand-off distance of the bubble centre from the wall and the crevice size were considered.
Changing these parameters significantly alters the behaviour of the bubble collapse and its associated impact on the wall.

For smooth-wall configurations the final collapse of the bubble results in the maximum wall-pressure, rather than the liquid jet that impinges it.
This is in agreement with experimental studies.
A similar behaviour is observed for smaller crevice sizes, albeit for larger crevices the jet-induced pressures are more significant than the collapse pressures. 
The presence of the crevice results in a complex collapse process. Reflection and wave superposition result in wave dynamics, which can induce significant post-collapse pressures.

The part of the bubble interface initially in contact with the high-pressure liquid plays an important role in the collapse behaviour. 
The bubble collapse behaviour was qualitatively similar for the smooth-wall and small-crevice cases, since the pressure distribution at the interface was comparable.
However, large crevices led to a significantly different bubble-liquid interface area, and thus qualitatively different dynamics.
The effect of the wall geometry on the collapse behaviour and wall pressure was smaller for wall-detached cases. 

Lastly, we considered the potential for cavitation erosion.
Pressures were recorded over a larger part of the wall. 
The presence of the small crevice leads to a significant pressure over the entire crevice bottom, as opposed to the smooth-wall cases when largest pressures occurred at the wall centre.
For all rough configurations, high pressures also occur at the crevice edges, where they induce stresses that can result in material damage. 
The pressure impulse also increased by about $50\%$ from the smooth-wall to small-crevice case, indicating an increased potential for material damage. 

While assessing the effects of surface topology on hydrodynamics is a necessary step towards understanding this complex process, prediction of actual cavitation erosion also requires investigations of exposed materials. 
Coupled fluid-material simulations that incorporate suitable material models, and thus also represent elastic and plastic deformation, are one way to accomplish such investigations.

\section*{Acknowledgments}

The research stay of T.T. at Caltech was supported by the Deutscher Akademischer Austauschdienst (DAAD), the TUM Graduate School, and the ERC Advanced Grant NANOSHOCK (2015).
S.H.B., K.S., and T.C. acknowledge support from the US Office of Naval Research under grant numbers N0014-18-1-2625 and N0014-17-1-2676.

\section*{Declaration of Interests} 

The authors report no conflict of interest.

\section*{Supplementary videos}
Supplementary videos of the reported simulations are available online. 

Videos of an air bubble collapsing onto a wall. The file names correspond to the different configurations considered. 
The videos show numerical schlieren (left) and log-scale pressure field (right). Gas volume fraction $\alpha_g$ is shown as a shaded area of decreasing opacity with decreasing $\alpha_g$ (left), while the $\alpha_g = 0.5$ bubble interface is shown as a solid curve (right). Time and pressure correspond to a $R_0 = \SI{400}{\micro\meter}$ bubble that is exposed to a driving pressure of $p_\infty=\SI{e7}{\pascal}$.

\vspace{2cm}
\bibliographystyle{jfm}
\bibliography{paper}

\end{document}